\documentclass[12pt]{article}
\usepackage{amssymb,amsmath}
\usepackage{hyperref}
\usepackage{cite}
\usepackage{mciteplus}
\def\beq{\begin{equation}}
\def\eeq{\end{equation}}
\def    \bea    {\begin{eqnarray}}
\def    \eea    {\end{eqnarray}}
\def \ba{\begin{align}}
\def \ea{\end{align}}

\def\non{\nonumber}
\newcommand{\Mb}{{\bar{M}}}

\newcommand{\Z}{\mathbb{Z}}
\newcommand{\R}{\mathbb{R}}

\renewcommand{\a}{\alpha}
\renewcommand{\b}{\beta} 

\def\dt{\delta}

\newcommand{\dl}{\delta}

\newcommand{\ta}{\tau}
\renewcommand{\t}{\theta}

\def\m{\mu}
\def\n{\nu}
\renewcommand{\H}{\mathcal{H}}

\def\L{\mathcal L}
\def\K{\mathcal K}
\def\V{\mathcal V}
\def\R{\mathcal R}
\newcommand{\X}{\mathbb{X}}
\renewcommand{\d}{\partial}

\providecommand{\openone}{\leavevmode\hbox{\small1\kern-3.8pt\normalsize1}}
\def\xt{\tilde{X}}

\def\dt{\tilde{\d}}
\def\pt{\tilde{P}}

\numberwithin{equation}{section}
\begin{document}

\begin{titlepage}
%\rightline\today 

\begin{center}
\vskip 2.5cm
{\Large \bf {
A Double Sigma Model for \\Double Field Theory}}\\
\vskip 2.0cm
{\large {Neil B. Copland}}
\vskip 0.5cm
{\it {Centre for Quantum Spacetime}}\\
{\it {Sogang University}}\\
{\it {Seoul 121-742, Korea}}\\
ncopland@sogang.ac.kr

\vskip 2.5cm
{\bf Abstract}
\end{center}

\vskip 0.5cm

\noindent
\begin{narrower}
We show that generalised metric equation of motion of doubled field theory (the vanishing of the generalised Ricci tensor) can be derived as the background field equation of a double sigma model. Thus the double field theory is the effective field theory for the sigma model.
\end{narrower}

\end{titlepage}

\section{Introduction}

String theory on a $d$-torus possess $O(d,d)$ T-duality invariance. Although this is particular to strings, rather than point particles (the duality mixes momentum in the compact directions with string winding around them), the $O(d,d)$ symmetry can still be observed in the effective field theory when it is dimensionally reduced on a torus. In double field theory\cite{Hull:2009mi} describing $D$ dimensional physics an additional set of co-ordinates are introduced leading to a manifest $O(D,D)$ symmetry. The dual co-ordinates associated to toroidal dimensions are conjugate to the winding in those directions (just as the original co-ordinates are conjugate to momentum), and the $O(d,d)$ symmetry in these directions should be restricted to $O(d,d;\Z)$ to preserve periodic boundary conditions. These $O(d,d;\Z)$ transforms can be thought of as T-dualities relating equivalent string backgrounds when reduced to the undoubled effective field theory in terms of the ordinary string massless fields.

However, any other, non-compact, dimensions are (usually) doubled as well. The dual co-ordinates in these directions no longer have an interpretation as winding and only in one (the canonical) duality frame can we express the double field theory fields in terms of undoubled fields representing a physical string background. Nevertheless, compact and non-compact directions are formally identical and we can describe the whole theory in terms of the generalised metric, $\H$, and dilaton, $d$, where the generalised metric indices range over the  doubled set of co-ordinates. The doubled field theory fields $\H$ and $d$ combine the undoubled metric, $g$, and anti-symmetric tensor, $b$, along with the ordinary dilaton, $\phi$, and the structure and geometry of double space becomes clearer when double field theory is expressed in terms of them. The double field theory equation of motion is expressed as the vanishing of the generalised Ricci tensor $\R_{MN}$\cite{Hohm:2010pp} (a function of $\H$ and $d$) and recent work has focussed on finding a doubled differential geometry of $\H$ in which $\R_{MN}$ arises naturally \cite{Hohm:2010xe,Jeon:2011cn} beyond the ordinary Riemannian geometry of $d$.

In \cite{Copland:2011yh} it was shown that in the doubled formalism (a sigma model in which $d$-dimensional toroidal fibres which are also doubled giving manifest $O(d,d)$ symmetry) this generalised Ricci tensor also occurs, albeit in a reduced form. It is reduced in the sense that, since in the doubled formalism only the fibre is doubled and none of the fields depend on these doubled co-ordinates, we should restrict the double field theory background on which $\R_{MN}$ is defined to be of this form. The vanishing of this reduced generalised Ricci tensor is the background field equation of the doubled formalism. One is immediately led to ask whether a similar correspondence exists for the full range of double field theory backgrounds. This of course demands defining a double sigma model whose target space is a general double field theory background. 

The main result of this paper is to show that there is such a doubled sigma model whose  one-loop background field equation is the vanishing of the full double field theory Ricci tensor,  implying that double field theory is the effective theory for the sigma model and a two-loop calculation could yield higher order corrections. The sigma model is the straightforward extension of the doubled formalism action without manifest Lorentz invariance considered in \cite{Berman:2007xn} to the case where all co-ordinates are doubled with the fields able to depend on this enlarged set (such actions were considered in \cite{Tseytlin:1990va,Tseytlin:1990nb}). The crucial extra feature here is that we impose the strong constraint that the double field theory is required to obey on the sigma model background as well. This means that half of the components of the equation of motion can be integrated, which is exactly as needed to demonstrate classical Lorentz invariance, show invariance under the double gauge transforms of double field theory and derive the background field equation. 

%%%%%%%%%%
These results are of interest from various perspectives. As remarked in \cite{Berman:2007xn} the generalisation of the doubled formalism to allowing the generalised metric to depend on the fibre co-ordinates was unknown, especially the nature and role of the chirality constrain. Here we establish the non-manifestly Lorentz invariant form of that action, with  the level-matching constraint of double field theory allowing the generalisation. Only half of the components of the chirality constraint are imposed. Although closely related to earlier models\cite{Berman:2007xn,Sfetsos:2009vt,Avramis:2009xi}, the sigma model is $O(D,D)$ invariant, including the dependence on the doubled co-ordinates (via the level-matching constraint). Given the classical equivalence of the sigma model to the standard sting,  it is not necessarily obvious that it should be quantum equivalent. This is established here at the one-loop level for a completely general $g$, $b$ and $\phi$ background, indirectly through the relation to double field theory which we know reproduces the equations of motion following the effective action for the ordinary string.

However what has really prompted this investigation is that the sigma model can give a new perspective on double field theory. Since the one loop background field expansion gives us the known two derivative double field theory, the two loop calculation will give higher order corrections in $O(D,D)$ invariant form. Though there has been a huge amount of recent interest in double field theory, the form of the higher order corrections is an open question\cite{Hohm:2011si}, though the work of Meissner\cite{Meissner:1996sa} on the dimensional reduced theory may provide guidance. The calculation of these corrections is well underway. Another topic of interest is the nature of the generalised $O(D,D)$ of double field theory which also underlies the sigma model. Here we confirm the importance of the ``generalised Ricci tensor'' whose vanishing is the background field equation of motion. It combines the $g$and $b$ equations in an $O(D,D)$ symmetric manner. A natural extension to this paper is finding the correct doubled background field expansion to simplify the calculation which would require (and guide) the determination of doubled differential geometrical versions of concepts like geodesics and normal co-ordinates for the 2D-dimensional space (see discussion in the final section). 

Finally, having established the connection of the sigma model to the generalised metric formulation of double field theory with the strong constraint enforced,  one could relax the constraint and allow more general dependence on the doubled co-ordinates (i.e. some form of true doubling, closer to the original goal of double field theory). Recent work has shown that  when compactifying double field theory to obtain gauged supergravities in 4 dimensions, the strong constraint of double field theory gives a constraint in 4 dimensions that is too strict to give the most general gaugings\cite{Geissbuhler:2011mx,Aldazabal:2011nj}. Indeed it has been shown that the strong, and even the weak, constraint can be relaxed slightly whilst maintaining the gauge symmetry of the double field theory action \cite{Hohm:2011cp,Grana:2012rr}. It would be interesting to compare the more general constraints on gauge invariance in the fiedl theory with those for a consistent gauge invariant sigma model if one can be found. This would lead to a geometric worldsheet description of a wider range of non-geometric backgrounds, echoing the original motivations of the doubled formalism. 
%%%%%%

We begin with a brief introduction to the generalised metric formulation of doubled field theory\cite{Hohm:2010pp} and its relation to the doubled formalism, extending the discussion of the strong form of the level-matching constraint to the sigma model case.  In Section 3 we introduce the general chiral sigma model with which we will work, examining its equations of motion in the presence of the strong constraint, which can then be used to confirm its classical Lorentz invariance and invariance under double gauge transformations after the inclusion of a topological term familiar from the doubled formalism. Section 4 is concerned with background field expanding this action to obtain the condition for Weyl invariance at the one-loop level and relating this to double field theory, including necessary dilaton terms.  In the last section we present some conclusions and discussion.

\section{Double Field Theory}

In this section we review the pertinent points of double field theory, as formulated by Hull, Zwiebach and Hohm\cite{Hull:2009mi, Hull:2009zb,Hohm:2010jy, Hohm:2010pp}.  The fundamentals of doubled field theory were recently reviewed in \cite{Zwiebach:2011rg}. It is closely connected to the earlier works of Tseytlin and Siegel\cite{Tseytlin:1990va,Tseytlin:1990nb,Siegel:1993bj,*Siegel:1993th,*Siegel:1993xq} and connection to Siegel's work is illustrated in \cite{Hohm:2010xe}. While the theory was originally meant to be truly doubled and go beyond the ordinary effective action we will be interested in the case where the `strong constraint' is enforced restricting the co-ordinate dependence of the fields. The theory can then be expressed in terms of the generalised metric, with new doubled geometrical structures coming to the fore with the theory possessing a new doubled gauge symmetry. However, the theory is no longer truly doubled and is equivalent to the effective action
\beq\label{EffAct}
S=\int dX\sqrt{-g}e^{-2\phi}\left[R+4(\d\phi)^2-\frac{1}{12}H^2\right]\, 
\eeq
for the metric, anti-symmetric two-form and dilaton, $g,b$ and $\phi$, which are functions of $D$ co-ordinates $X^i$. We introduce the dual co-ordinates $\xt_i$ which combine with the original co-ordinates to give a 2D vector $X^M=(\xt_i,X^i)$ which transforms in the fundamental of $O(D,D)$ and allow all the fields to depend on them. We then have corresponding derivatives $\d_M=(\tilde{\d}^i,\d_i)$ and the action can be written in terms of the generalised metric
\beq\label{Hhhz}
\H_{MN} = \left( \begin{array}{cc}
 g^{ij} & -g^{ik}b_{kj}\\
 b_{ik}g^{kj}& g_{ij} - b_{ik}g^{kl}b_{lj}
\end{array}\right)\, ,
\eeq
and the doubled dilation
\beq
e^{-2d}=\sqrt{g}e^{-2\phi}.
\eeq
The generalized metric transforms as an $O(D,D)$ tensor under and $O(D,D)$ rotation acting on the vector $X^M$ (to be contrasted with the complicated transformations of $g$ and $b$ individually) and is thus a natural object with which to build and $O(D,D)$ invariant action (the doubled dilaton is invariant under these transformations). The generalised metric has the property that raising the indices with the $O(D,D)$ invariant metric 
\beq\label{Lmetric}
L_{MN} = \left( \begin{array}{cc}
0 &\openone\\
\openone &0
\end{array}\right)
\eeq
gives its inverse, that is $\H^{-1}=L^{-1}\H L^{-1}$. In fact it is the most general matrix with this form, this form also indicating that it parameterises an $O(D,D)/(O(D)\times O(D))$ coset. In what follows $O(D,D)$ indices will always be raised and lowered with the metric $L_{MN}$.

The action of double field theory can be cast in Einstein-Hilbert form for a scalar function of the generalised metric and doubled dilaton which is denoted $\R$\cite{Hohm:2010pp}. The notation makes clear that in some sense the scalar is playing the r\^ole that the Ricci scalar does in ordinary gravity. The dilaton equation of motion is the vanishing of $\R$, whilst the variation of the action with respect to $\H^{AB}$  is proportional to $\K_{AB}\delta \H^{AB}$, where 
\begin{align}
\label{kis}
\begin{split}  
{\cal K} _{MN}={}&\frac{1}{8}\, \partial_{M}{\cal H}^{KL}
  \,\partial_{N}{\cal H}_{KL}
  -\frac{1}{4}(\partial_L - 2 (\partial_L d) ) 
  ({\cal H}^{LK} \partial_K {\cal H}_{MN})
 \\&+2 \,\partial_{M}\partial_N d\,  -\frac{1}{2} \partial_{(M}{\cal H}^{KL}\,\partial_{L}
  {\cal H}_{N)K}
  \\
  &+ \frac{1}{2} (\partial_L - 2 (\partial_L d) )  \bigl({\cal H}^{KL} \partial_{(M}
   {\cal H}_{N)K}
  + {\cal H}^K{}_{(M}  \partial_K {\cal H}^L{}_{N)}  \bigr) \,.
   \end{split}
\end{align}
However, the field equation is not simply the vanishing of $\K_{MN}$, as the variation of $\H$ should preserve its coset form. In other words, since the original field satisfies $\H^{MN}L_{NK}\H^{KL}=L^{ML}$,   the field after variation $\H'=\H+\delta\H$ must satisfy the same relation. This constrains the form of the variation and thus the field equation is the vanishing of
\begin{align}
\R_{MN}={}&\frac{1}{2}\left(\K_{MN}-\H^{\ P}_{M}\K_{PQ}\H^{Q}_{\ N}\right)\, .
\end{align}
As with the scalar, $\R_{MN}$ is playing the r\^ole of the Ricci tensor, and for want of a better name we shall refer to it as the generalised Ricci tensor (note that it is {\it not} the Ricci tensor constructed from the generalised metric $\H$ and it does not simply trace to give the generalised Ricci scalar).  The ordinary Ricci tensor encodes the differential geometry associated to the metric $g$, whereas here the generalised Ricci tensor somehow encodes the doubled $O(D,D)$ differential geometry associated to $g,b$ and $\phi$. Understanding this differential geometry from different perspectives has been the focus of much recent work\cite{Hohm:2010xe,Jeon:2010rw,Jeon:2011cn,Coimbra:2011nw}.

There has also recently been progress towards a full doubled supergravity\cite{Jeon:2011vx,Hohm:2011zr,*Hohm:2011dv}(see also \cite{Coimbra:2011nw}) and the development of a closely related generalised geometry for M-theory \cite{Berman:2010is,*Berman:2011kg,*Berman:2011pe,*Berman:2011cg, Thompson:2011uw}, while additional interesting extensions can be found in \cite{Jeon:2011kp,*Hohm:2011ex,*Jensen:2011jn,*Albertsson:2011ux,Hohm:2011cp}. 

\subsection{The level-matching constraint}\label{sec:const}

A crucial ingredient of doubled field theory is the level matching constraint which comes in both weak and strong forms. It can be expressed in terms of a differential operator 
\beq
\Delta=L^{MN}\d_M\d_N=2\d_i\dt^i\, .
\eeq
The weak form of the constraint comes from closed string level matching in a toroidal background and is that this operator should annihilate the fields and gauge parameters of the theory. The strong form of the constraint was required in writing a background independent form of the double field theory action and is is that the operator should also annihilate products of fields, in particular this implies
\beq\label{strongAB}
\d_M A\, \d^M B=0
\eeq
for any fields or gauge parameters $A, B$ in the theory. Since later we will be interested in this constraint in the sigma model context let us examine it in more detail as in \cite{Hohm:2010jy}. First consider a field which is a single Fourier mode
\beq
A(\xt_i, X^i)=Ae^{i(\pt^i\xt_i+P_iX^i)}\, .
\eeq
The momentum $P_M$ also transforms in the fundamental of $O(D,D)$ and the constraint tells us that it is null as an $O(D,D)$ vector,
\beq
P\cdot P:=L^{MN}P_MP_N=0\, .
\eeq
This is true for all fourier modes of all fields, and the strong form of the constraint (\ref{strongAB}) tells us that for any two momentum vectors $P, P'$ associated to different Fourier components of any field in the theory
\beq
P\cdot P'=0\, ,
\eeq
i.e. they are null and orthogonal which means that all the momentum vectors lie in an isotropic subspace of ${\mathbb R}^{2D}$. The canonical choice for such a subspace would be that of all $P^M$ with $\pt^i=0$. 
If this is the case the fields only depend on the co-ordinates of a totally null $D$-dimensional subspace of the double space. The canonical choice corresponding to $\pt^i=0$ is that with no dependence on the dual co-ordinates. Now any two maximal ($D$-dimensional here) isotropic subspaces are dual to each other under an $O(D,D)$ rotation\footnote{We recall that in any compact directions we should restrict to $O(d,d;{\mathbb Z})$ to preserve the periodic boundary conditions.}, so we can always to rotate to the canonical one. 

We will also be interested in the constraint in the sigma model context. With this in mind the picture given here makes clear how strong the strong constraint is: if the $X$'s are target space fields the strong constraint will hold even when evaluated at different points on the world sheet. That is
\beq\label{ABsigma}
L^{MN}\frac{\dl A(X(\sigma))}{\dl X^M(\sigma)}\,\frac{\dl B(X(\sigma'))}{\dl X^N(\sigma')}=0\, ,
\eeq
%\beq\label{ABsigma}
%\d_M A(X(\sigma))\, \d^M B(X(\sigma'))=0\, ,
%\eeqL
because thinking of $A$ and $B$ as Fourier transforms of functions of momentum, differentiation at any point will bring down a the momentum vectors in the same null subspace, and the contraction with $L$ will always give zero.

 \subsection{Relation to the doubled formalism}

The doubled formalism\cite{Hull:2004in,Dabholkar:2005ve,*Hull:2009sg,Hull:2006va,Hull:2006qs} was introduced to describe string backgrounds with a toroidal fibre in a manner which makes the T-duality action on the fibre into a manifest symmetry of a sigma model. The fibres are doubled and in this new partially doubled target space, backgrounds such as T-folds and twisted tori\cite{Hull:2004in,Dabholkar:2005ve,Hull:2005hk,Shelton:2006fd,*Shelton:2005cf,*Becker:2006ks} would have a geometrical description: in a T-fold T-duality transition functions are allowed, resulting in non-geometric backgrounds which are not manifolds.

The action is a sum of base and fibre parts, with the base part given by an ordinary sigma model action and the fibre part of the action is a sigma model action for the generalised metric $\H$ of the doubled fibres. Crucially $\H$ only depends on the base co-ordinates and there is isometry in all the fibre directions. The action must be supplemented by a chirality constraint that ensures the fibre directions can be thought of as chiral bosons on the worldsheet so that the doubling does not increase the degrees of freedom. The constraint can be incorporated into the action\cite{Berman:2007xn} and the resulting action on the fibre lacks manifest Lorentz invariance and takes a form previously considered by Tseytlin\cite{Tseytlin:1990va}, 
\beq\label{ODaction}
S_{fib}= \frac{1}{2} \int d^2\sigma\left[ -\H_{MN}(Y) \partial_1 X^M \partial_1 X^N + L_{MN} \partial_1 X^M \partial_0 X^N \right]\,   ,
\eeq
where $X^M$ are the doubled set of fibre co-ordinates and $Y$ represents the base co-ordinates.
%\beq
%\label{eqConstraint}
%d\X^M = L^{MN}\H_{NP}\ast d\X^P .
%\eeq
An important point here is the second-order equation of motion for the fibre co-ordinates
\bea
\d_1 \left( \H_{MN} \d_1 \X^N \right) = L_{MN} \d_1\d_0 \X^N,
\eea
 can be integrated (the gauge invariance of the action under $\X^A\rightarrow \X^A+f(\tau)$ is then used to remove an integration function of $\ta$) to give the chirality constraint which was incorporated into the action. Once the action is in this form we can perform a background field expansion to test the quantum consistency of the doubled formalism. This test is one loop in $\alpha'$ and compliments the one string-loop calculation in\cite{Berman:2007vi} (other quantum tests were performed in \cite{HackettJones:2006bp,Chowdhury:2007ba}). In order to have UV finiteness and worldsheet Weyl invariance  mainained at 1-loop a certain tensor $W_{\a\b}$, constructed from $\H_{MN}$ and $g_{ab}$ and their derivatives, must vanish (to compare, for an ordinary sigma model of only the metric $g$ the condition would be the vanishing of the Ricci tensor) and this is equivalent to the beta-functional equations for the ordinary string sigma model on the same fibred background\cite{Berman:2007xn}.

It was shown in \cite{Copland:2011yh} that $W_{\a\b}$ is proportional to $\R_{MN}$ of doubled field theory when the doubled field theory is restricted to a fibered background of the type that the doubled formalism describes. One then wonders whether the sigma model with doubled fibre could be extended to a wholly doubled sigma model, one whose background field equation is the same as the double field theory equation of motion, $\R_{MN}=0$, on any background that doubled field theory can describe.
%%%%%%%%%%
\section{Double sigma model}

We propose that the sigma model whose background field expansion gives double field theory is given by
\beq\label{NDaction}
S= \frac{1}{2} \int d^2\sigma\left[ -\H_{MN} \partial_1 X^M \partial_1 X^N + L_{MN} \partial_1 X^M \partial_0 X^N \right]\,   ,
\eeq
where $X^M$ is a doubled co-ordinate $X^M=(\xt_i,X^i)$ and $\H=\H(X^M)$ can depend on any of these doubled co-ordinates, subject to the strong form of the level-matching constraint. As before, $L_{MN}$ is the $O(D,D)$ invariant constant metric (\ref{Lmetric}) and we observe that this action the straightforward generalisation of the doubled formalism action without manifest Lorentz invariance considered in \cite{Berman:2007xn}, (\ref{ODaction}), without a base-fibre split. Such actions were considered by Tseytlin in \cite{Tseytlin:1990va,Tseytlin:1990nb} and it was noted that in the case where generalised metric only depends on the $X^i$ co-ordinates it can be derived from the first-order form of the ordinary sigma-model upon putting $p_i=\d_1\xt_i$. Here we allow the generalised metric to also depend on the dual co-ordinates subject to the strong constraint. As explained in Section \ref{sec:const} this means we can always perform an $O(D,D)$ rotations so that the fields only depend on $X^i$ and as all $O(D,D)$ indices in the action are contacted properly its form is preserved under such a rotation. While things can normally be shown more clearly in this frame, in what follows it is not necessary to break the $O(D,D)$ invariance in order to use the equation of motion along with the strong constraint to derive what we want.

The equation of motion following from this action is
\beq\label{eom}
\d_1(\H_{MN}\d_1X^B-L_{MN}\d_0X^N)=\frac{1}{2}\d_M\H_{NP}\d_1X^N\d_1X^P\, ,
\eeq
which could also be written
\beq
D_1(\H_{MN}\d_1X^B)=L_{MN}\d_1\d_0X^N\, ,
\eeq
where $D$ is a covariant derivative constructed with the Levi-Civita connection for $\H_{MN}$, in this case pulled back to the worldsheet. In the fibred case we had a total derivative, but here we proceed and integrate anyway giving
\beq\label{eomint}
\H_{MN}\d_1X^N-L_{MN}\d_0X^N=\frac{1}{2}\int d\sigma'_1\epsilon(\sigma_1-\sigma'_1)[\d_M\H_{NP}\d_1X^N\d_1X^P](\sigma')\, ,
\eeq
using $\epsilon(\sigma)=(\theta(\sigma)-\theta(-\sigma))/2$ so that $\d_{\sigma}\epsilon(\sigma)=\delta(\sigma)$. The notation under the integral is intended to indicate that all $X$'s are functions of $\sigma_1'$. Here we have set to zero a function of $\sigma_0$ introduced by the integration by assuming the boundary conditions allow us to do so (this was done in similar circumstances in \cite{Tseytlin:1990nb,Tseytlin:1990va}). In the doubled formalism there was a gauge invariance in the action that allowed us to set such functions to zero, we have not yet been able to find a similar symmetry here, but we note that the consistency of the final answer as evidence that setting the function to zero is the correct thing to do.

When we are required to use the integrated equation of motion the free index will always be contracted with an $X^M$ derivative so that the strong level-matching constraint will imply the vanishing of the non-local term through \eqref{ABsigma}. To understand this  we rotate to the canonical duality frame using $O(D,D)$, so that all fields are functions of the $X^i$ co-ordinates only and $\dt^i \H_{MN}=0$. We can safely integrate the  components of the equation of motion with an upper $i$ index giving\footnote{Curiously, since $\H$ does not depend on $\xt_i$ we could shift $\xt_i\rightarrow\xt_i+f_i(\sigma_0)$ to remove a possible function of $\sigma_0$, but this is the wrong half of the components to remove the integration constant in this equation. As before we rely on boundary conditions to do so.}
\beq\label{eomdual}
\H^{i}_{\ N}\d_1X^N-L^{i}_{\ N}\d_0X^N=0\, .
\eeq
We find although we can only integrate half the components of the equation of motion, it is only these components that we need in first order form to perform any necessary manipulations. Substituting \eqref{eomdual} into the remaining components of the equation of motion one can reproduce the equations of motion of the ordinary sigma model
\beq
\L_s=\frac{1}{2}g_{ij}\d_\mu X^i\d^\mu X^j +\frac{1}{2}\epsilon^{\m\n}b_{ij}\d_\mu X^i\d_\n X^J\, ,
\eeq
as we expect given the equivalence to the phase space form of the action once in the canonical duality frame.

\subsection{Lorentz invariance}

This action is not manifestly Lorentz invariant on the worldsheet and we check that it is actually possesses this symmetry by examining its variation under the Lorentz transform $\delta_L X^M=-\sigma_0\d_1X^M-\sigma_1\d_0X^M$. Up to total derivatives the variation is given by
\begin{align}
\delta_L S&=-\frac{1}{2}\int d^2\sigma (-L_{MN}( \d_0 X^M\d^N_0 X+ \d_1 X^M\d_1 X^N)+2H_{MN} \d_0 X^M\d_1 X^N)\\
&=-\frac{1}{2}\int d^2\sigma(L_{MN} \d_0 X^N-\H_{MN} \d_1 X^N)L^{MK}(L_{KL} \d_0 X^L-\H_{KL} \d_1 X^L)\, .\label{Lorentzvar}
\end{align}
We can see this vanishes on the equations of motion either by using (\ref{eomint}) to get 
\begin{align}
\delta_L S={}&-\frac{1}{8}\int d^2\sigma\int d\sigma'_1\epsilon(\sigma_1-\sigma'_1)\left[\d_M\H_{NP}\d_1X^N\d_1X^P\right](\sigma')\non\\
&\times\int d\sigma''_1\epsilon(\sigma_1-\sigma''_1))\left[\d^M\H_{KL}\d_1X^K\d_1X^L\right](\sigma'')\, ,
\end{align}
which is zero by the level-matching constraint \eqref{ABsigma}\footnote{This means rather than thinking of the condition that $\H_{MN}$ obey the strong constraint as being imposed externally, it could be thought of as following from the condition of classical Lorentz invariance.}. Alternatively we could choose to examine the $O(D,D)$ invariant expression \eqref{Lorentzvar} in the canonical frame where the integrated half of the equation of motion (\ref{eomdual}) shows that the dual components of the bracketed expression vanish. The off-diagonal nature of $L$ means the whole expression is zero and the action is classically Lorentz invariant. Lorentz invariance of the canonical frame sigma model was noted in \cite{Tseytlin:1990va}, under local-Lorentz rotations of a worldsheet 2-bein which can be repeated here, while Lorentz invariance of similar actions was also considered in\cite[Section 5.6]{Thompson:2010sr}). Those works showed that chiral sigma models can have a modified Lorentz symmetry which reduces to the ordinary Lorentz symmetry on shell. Such a symmetry may exist here.

\subsection{Double gauge transformations and the topological term}

Recall that doubled field theory posses a gauge symmetry under which the transformation of the generalised metric $\H$ with parameter $\xi^M$ is
\beq
\delta_{\xi}\H_{MN}=\xi^P\d_P\H_{MN}+(\d_M\xi^P-\d^P\xi_M)\H_{PN}+(\d_N\xi^P-\d^P\xi_N)\H_{MP}\, ,
\eeq
which can be considered as acting as a generalised Lie derivative\cite{Hohm:2010pp} (compare with the transformation of the metric by a Lie derivative under diffeomorphisms).
A vector should transform as
\beq
\delta_\xi V^M=\xi^P\d_PV^M+(\d^M\xi_P-\d_P\xi^M)V^P\, .
\eeq
This implies that in the canonical duality frame, where nothing depends on $\xt$, the components transform as
\bea
\delta_\xi V^i&=&\xi^k\d_kV^i-\d_k\xi^iV^k\, ,\label{trans1}\\
\delta_\xi \tilde{V}_i&=&\xi^k\d_k\tilde{V}_i+\d_i\xi^k\tilde{V}_k+(\d_i\tilde{\xi}_k-\d_k\tilde{\xi}_i)V^k\, ,\label{trans2}
\eea
which we recognise as diffeomorphisms plus gauge transforms of $b$. Using the integrated components of the equation of motion we have that
\beq\label{d1again}
\d_1\xt_i=g_{ij}\d_0X^j+b_{ij}\d_1X^j\, ,
\eeq
and when the gauge parameters only depend on $X^i$ the metric and anti-symmetric tensor transform as\cite{Hohm:2010pp}
\bea
\delta_{\xi}g_{ij}&=&\L_{\xi}g_{ij}\, ,\\
\delta_{\xi}b_{ij}&=&\L_{\xi}b_{ij}+\d_i\tilde{\xi}_j-\d_j\tilde{\xi}_i\, ,
\eea
where
\beq
\L_{\xi}t_{ij}=\xi^k\d_kt_{ij}+\d_i\xi^kt_{kj}+\d_j\xi^kt_{ik}
\eeq
for general tensor $t_{ij}$. The vector $\d_1X^M$ thus transforms as
\bea
\delta_\xi(\d_1X^i)&=&-\d_k\xi^i\d_1X^k\, ,\label{d1xvar}\\
\delta_\xi(\d_1\xt_i)&=&(\xi^k\d_kg_{ij}+\d_i\xi^kg_{kj})\d_0X^j+(\xi^k\d_kb_{ij}+\d_i\xi^kb_{kj})\d_1X^j\nonumber\\
&&+(\d_i\tilde{\xi}_j-\d_j\tilde{\xi}_i)\d_1X^j\\
&=&\xi^k\d_k\d_1\xt_i+\d_i\xi^k\d_1\xt_k+(\d_i\tilde{\xi}_j-\d_j\tilde{\xi}_i)\d_1\xt^j\, ,\label{d1xtvar}
\eea
which shows that $\d_1X^M$ transforms as (\ref{trans1}) and (\ref{trans2}), the correct manner for an $O(D,D)$ vector, so we can conclude that the transformation of the  $\H_{MN}\d_1X^M\d_1X^N$ term in the double sigma model is only a transport term, the correct behaviour for this part of the action to have double gauge symmetry.

When we examine the $L_{MN}$ term in the action we encounter a problem as $\d_0 X^M$ does not have a simple gauge transformation, more specifically $\d_0\xt_i$ does not. By using \eqref{d1again} in the other half of integrated equation of motion we see
\beq
\d_0\xt_i=b_{ij}\d_0X^j+g_{ij}\d_1X^j-\frac{1}{2}\int d^2\sigma\epsilon(\sigma-\sigma')\left[\d_i\H_{MN}\d_1X^M\d_1X^N\right](\sigma').
\eeq
The first two terms together transform correctly as the dual components of a vector, but the non-local term does not have a simple gauge transformation. We can remove $\d_0\xt_i$ from the action by performing an integration by parts, shifting
\begin{align}\label{Lptd}
L_{MN}\d_1X^M\d_0X^N=\d_1X^i\d_0\xt_i+\d_1\xt_i\d_0X_i\rightarrow2\d_1\xt_i\d_0X_i\, .
\end{align}
To evaluate the gauge variation we can use the variation of $\d_1\xt_i$ from (\ref{d1xtvar}), while $\d_0X^i$ transforms similarly to (\ref{d1xvar}). We find
\begin{align}
\delta_\xi \left(L_{MN}\d_1X^M\d_0X^N\right)={}&2\left(\xi^k\d_k\d_1\xt_i+\d_i\xi^k\d_1\xt_k+(\d_i\tilde{\xi}_j-\d_j\tilde{\xi}_i)\d_1\xt^j\right)\d_0X^i\non\\
&-2\d_1\xt_i \d_k \xi^i\d_0X^k+t.d.\\
={}&\xi^k\d_k\left(2\d_1\xt_i\d_0X^i\right)+t.d.\, ,
\end{align}
where the $\tilde{\xi}$ terms also give a contribution to the total derivative (t.d.). The variation reduces to the transport term of $2\d_1\xt_i\d_0X_i$, but the difference between this transport term and the transport term of the original $L_{MN}$ term in the action is not a total derivative and it seems we do not have gauge invariance. However, we recall that a topological term for the doubled formalism was introduced in \cite{Hull:2006va}, where is was needed to maintain invariance under large gauge transformations while imposing the chirality constraint by gauging an associated current,  and also found necessary in \cite{Berman:2007vi} to show equivalence of the torus partition function to the undoubled case. The equivalent term here (in the canonical duality frame) is given by
\beq\
S= \frac{1}{2} \int d^2\sigma\, \Omega_{MN} \partial_1 X^M \partial_0 X^N\,   ,
\eeq
where
\beq\label{omega}
\Omega_{MN} = \left( \begin{array}{cc}
 0& 1\\
-1& 0
\end{array}\right)\, ,
\eeq
and introducing it here provides exactly the total derivative we added in \eqref{Lptd}. We see that under gauge transformations
\beq
\delta_\xi \left((L_{MN}+\Omega_{MN})\d_1X^M\d_0X^N\right)=\xi^k\d_k\left((L_{MN}+\Omega_{MN})\d_1X^M\d_0X^N\right)+t.d.\, ,
\eeq
so that the total action
\beq\label{NNDaction}
S= \frac{1}{2} \int d^2\sigma\left[ -\H_{MN} \partial_1 X^M \partial_1 X^N + L_{MN} \partial_1 X^M \partial_0 X^N+ \Omega_{MN} \partial_1 X^M \partial_0 X^N \right]\, 
\eeq
is invariant under double gauge transformations. The topological term is Lorentz invariant and does not affect the equations of motion since it is a total derivative, so it does not affect the conclusions above and will play no r\^ole in the background field expansion. We see the effect of the topological terms is to remove $\d_0\xt_i$ from the action.

The tensor $\Omega_{MN}$ is not invariant under $O(D,D)$ transformations, but rather for co-ordinates related to the canonical ones by $X'^M=h^M_{\ N}X^N$ we have $\Omega'_{MN}=h^P_{\ M}h^Q_{\ N}\Omega_{PQ}$. For example, under the $O(D,D)$ transform which exchanges the canonical co-ordinates with their duals  $\Omega_{MN}$ would change sign giving
\beq
\L_{top}=-\frac{1}{2}\left(\d_1\xt_i'\d_0X'_i-\d_1{X^i}'\d_0\xt'_i\right)=\frac{1}{2}\left(\d_1\xt_i\d_0X_i-\d_1{X^i}\d_0\xt_i\right)\, .
\eeq
This ensures that $\d_0\xt_i=\d_0{X^i}'$ is still eliminated from the action, as it is still given by the components of the equation of motion that can not be integrated. If the topological term was invariant then $\d_0\xt'_i$ would be eliminated wrongly instead. This probably reflects that the sigma model should come from imposing a chirality constraint which has only $D$ components and is not $O(D,D)$ invariant (compare also with how only half the components of the current are gauged in \cite{Hull:2006va} where the topological term was introduced). While we would prefer to maintain the $O(D,D)$ invariance of the action we reiterate that the topological term is only required here to show double gauge invariance, and is not involved in what follows. 

In conclusion the chiral sigma model \eqref{NNDaction} subject to the strong level-matching constraint is classically lorentz invariant, invariant under double gauge transforms and equivalent to the ordinary string sigma model.

%%%%%%%%%%%%
\section{Background field expansion}

In order to examine the quantum behaviour of the sigma model we perform a background field expansion in quantum fluctuations around a classical background. Expanding to second order in fluctuations will allow us to calculate the one-loop background field equations. These equations give the necessary conditions on the background in order that worldsheet Weyl invariance be preserved, as well as guaranteeing freedom from UV divergences (at this order). The calculation simultaneously provides a check that worldsheet Lorentz invariance is also preserved at this order. It is through these background field equations that the connection to double field theory will be made.

We begin by writing the field $X^M$ as a classical background piece (which solves the classical equations of motion) plus a quantum fluctuation, $X^M=X^M_{cl}+\pi^M$. While one can expand in whatever one likes, expansion in such a fluctuation will not in general be covariant. Instead we expand in $\xi^M$ defined as the tangent vector to the geodesic from $X^M_{cl}$ to $X^M_{cl}+\pi^M$ whose length is equal to that of the geodesic\cite{Honerkamp:1971sh,AlvarezGaume:1981hn}. Since this is a contravariant vector, the expansion will naturally be organized in terms of covariant objects. This technique was first applied to chiral boson models of the doubled formalism in \cite{Berman:2007xn} and extended in \cite{Berman:2007yf,Copland:2011yh}. We follow these works in using the algorithmic method of performing the expansion described in \cite{Mukhi:1985vy}.

The background field expansion for general chiral sigma models was  considered in \cite{Avramis:2009xi} and general expressions for the Weyl and Lorentz anomaly terms were written down, before specialising to a specific class of backgrounds with group structure. In \cite{Sfetsos:2009vt} such group manifolds were also considered and both cases the end result was the vanishing of a tensor constructed from the structure constants.

\subsection{Expanding the doubled sigma model}

We perform the background field expansion starting from the Lagrangian of (\ref{NDaction}) without topological term. At first order the terms vanish due to the background being a solution of the equations of motion, as should always be the case. At second order in $\xi^M$ we get
\begin{align}\label{HLlag}
2{\mathcal L}_{(2)}={}&-\H_{MN}D_1\xi^M D_1\xi^N+L_{MN}D_0\xi^M D_1\xi^N\non\\
&-R_{KMNL}\xi^M\xi^N\d_1X^K \d_1X^L+L_{MN;K}\xi^K(D_0\xi^M\d_1X^N+\d_0X^M D_1\xi^N)\non\\
& +\frac{1}{2}D_M D_N L_{KL}\xi^M\xi^N\d_0X^{K}\d_1X^L\non\\
&+\frac{1}{2}\left( L_{KP}R^{P}_{\ MNL} +L_{LP}R^{P}_{\ MNK}\right)\xi^M\xi^N\d_0X^K\d_1X^L.\qquad
\end{align}
The next step is to obtain a propagator for the fluctuations, and then Wick contract all fluctuations out. In order to do this we introduce a vielbein which allows us to move to the chiral frame where the metrics are diagonal. This gives us kinetic terms for chiral and anti-chiral bosons in flat space, and since the propagator for these is known\cite{Tseytlin:1990nb} we can then Wick contract as intended. The effect of pulling the vielbeins through the covariant derivatives is to exchange the connection for the `spin-connection'
\beq
A^{M}_{\m\, N}=\Gamma^{M}_{\ \m N}+\d_\m\V{^M}_{\Mb}\V^{\Mb}_{\ N}\, .
\eeq
For the an undoubled string the connection terms can be dropped as there is a general argument why they cannot contribute to the Weyl divergence as they transform like a gauge field when pulled back to the worldsheet. In the doubled case the connection carries $O(D,D)$ rather than $O(D)$ indices and the arguments do not hold. In \cite{Copland:2011yh} it was explicitly demonstrated that in the doubled formalism theses terms do contribute to the Weyl divergence, so we must include them here. The new connection $A^{A}_{\m\, B}$ has the anti-symmetry
\bea\label{Hswap}
\H_{MK}A^{K}_{\m\, L}\H^{LN}&=&-A^{N}_{\m\, M}\, ,
\eea
wheras the vielbein piece $B^{M}_{\m\, N}=\d_\m\V{^M}_{\Mb}\V^{\Mb}_{\ N}=A^{M}_{\m\, N}-\Gamma^{M}_{\ \m N}$ obeys
\bea\label{Lswap}
L_{MK}B^{K}_{\m\, L}L^{LN}&=&-B^{N}_{\m\, M}\, .
\eea
The final complication is that in order to account for all one-loop contributions to the effective action, we must also include some terms at higher order in the expansion of the exponential of the effective action. Terms of the form $\xi\d\xi$ will only contribute to the logarithmic divergence at second order in this expansion. The necessary contractions  along with the propagator, when referred back to the original frame from the tangent frame, can be summarised as\cite{Berman:2007xn}
\begin{subequations}
\begin{align}
\langle \xi^{M}(z) \xi^{N}(z) \rangle 
&= \Delta_0\H^{MN}  +\theta L^{MN},\label{Wick1}\\
 \langle \xi^P \d_1 \xi^M \d_1 \xi^K \xi^Q \rangle &= -\Delta_0\left(\H^{M[Q} \H^{K]P} -L^{M[Q} L^{K]P}\right),\label{Wick2}\\
\langle \xi^P \d_1 \xi^M \d_0 \xi^K \xi^Q \rangle &=\Delta_0\left(\H^{M[Q} L^{K]P}+L^{M[Q} \H^{K]P}\right)+ 2 \t L^{M[Q} L^{K]P},   \label{Wick3}\\
\begin{split}\langle \xi^P \d_0 \xi^M \d_0 \xi^K \xi^Q \rangle &=\Delta_0\left(\H^{M[Q} \H^{K]P} +3L^{M[Q} L^{K]P}\right) \\&\qquad\qquad\qquad\ +  2\t \left( \H^{M[Q} L^{K]P}+L^{M[Q} \H^{K]P}\right),   \label{Wick4}\end{split}
\end{align}
\end{subequations}
The chiral and antichiral boson propagator integrals at zero distance combine into a divergent ordinary boson propagator integral, $\Delta_0$ and a parameter $\theta$, which keeps track of any violation of Lorentz invariance.

\subsection{The total divergence}

The complete contributions at one-loop proportional to $\Delta_0$ and $\theta$ can be written
\begin{subequations}
\begin{align}
\left(\R_{MN}+\left(\frac{1}{8}\d_P\H_{KL}\d_Q\H^{KL}+\d_L(\H^{KL}\d_P\H_{QK})\right)\H^P_{\ M}\H^{Q}_{\ N}\right)\quad\\
\times\d_1X^M\d_1X^N&\Delta_0\, ,\label{Delta10}\\
\d_L\left(\H^{KL}\d_B\H_{AL}\right)\H^{A}_{\ C}\,\d_1X^C\d_0X^B&\Delta_0\, ,\label{Delta00}\\
-\frac{1}{8}\d_M\H_{KL}\d_N\H^{KL}\d_0X^M\d_0X^N&\Delta_0\, ,\\
\H_{CK}\d_L\H^{KL}\left(-\d_M\H^C_{\ N}+\frac{1}{2}\d^C\H_{MN}\right)\d_1X^M\d_1X^N&\theta\, ,\label{theta11}\\
-\d_M\left(\H_{CK}\d_L\H^{KL}\right)\d_0X^M\d_1X^N&\theta\, ,\label{theta10}
\end{align}
\end{subequations}
We are heartened to see $\R_{MN}$ appear (for now we assume the doubled dilaton $d$ is constant, as we have not considered the terms in $\R_{MN}$ containing $d$ yet). Dealing with the $\theta$ terms first we can quickly see that after an integration by parts they are given by
\beq
\H_{CK}\d_L\H^{KL}\left(\d_0\d_1X^C-\d_1\H^C_{\ B}\d_1X^B+\frac{1}{2}\d^C\H_{AB}\d_1X^A\d_1X^B\right)\theta\, .
\eeq
this is proportional to the equation of motion (the unintegrated form) and so the $\theta$ terms drop out at one loop confirming the preservation of Lorentz invariance at this order. The $\Delta_0$ terms are slightly more subtle. The crucial point to observe is that in (\ref{Delta10}) and (\ref{Delta00}), $\d_0X^M$ is always contracted with a derivative. Thus we can use the integrated form of the equation of motion as the non-local terms will drop out (as always one could alternatively rotate to the canonical frame where it will only be necessary to use the half of the components of the equation of motion that can be integrated). Thus we can make the replacement of $\d_0X^M$ by $\H^M_{\ N}\d_1X^N$ yielding the final result
\beq\label{1loopresult}
{\mathcal L}_{eff}=\R_{MN}\d_1X^M\d_1X^N\Delta_0\, .
\eeq
\subsection{Dilaton}

We now turn to the terms of the generalised Ricci tensor which contain the dilaton, which we denote $R^{d}_{MN}$. This is the difference between what we have found so far from the background field expansion and the full $d$-dependant generalised Ricci tensor, and is given by
\bea\label{Rdil}
R^{d}_{MN}&=&\frac{1}{2}\d_L\H^{LK}\d_K\H_{MN}+\d_M\d_N d-\H_M^{\ K}\H_N^{\ L}\d_K\d_L d\non\\
&&-\d_Ld\left(\H^{KL}\d_M\H_{KN}+H^K_{\ (M}\d_K\H^L_{\ N)}\right)\, .
\eea
To $R^{d}_{MN}\d_1X^M\d_1X^N$ we can add zero in the form
\beq
\d_L d\, \H^{LK}\left(\d_1(\H_{KN}\d_1X^N-L_{KN}\d_0X^N)-\frac{1}{2}\d_K\H_{MN}\d_1X^M\d_1X^N\right)
\eeq
since the factor in brackets is the equation of motion which vanishes on shell. It cancels the first, second and fourth terms of leaving
\beq
\d_M(\H_N^{\ L}\d_L d)\d_0X^M\d_1X^N-\H_M^{\ K}\d_K(\H_N^{\ L}\d_L d)\d_1X^M\d_1X^N\, ,
\eeq
where in the first term we changed the order of differentiation and integrated by parts. Now the $\d_0X^M$ in the first term is contracted with a derivative, so  we can use the integrated equation of motion to eliminate it, cancelling the second term without introducing a non-local term. Thus $R^{d}_{MN}\d_1X^M\d_1X^N$ is a total derivative on shell and we are free to add it to the sigma model and we have correctly reproduced the whole generalised Ricci tensor multiplying $\d_1X^M\d_1X^N$.

A more complete understanding of the role of the doubled dilaton in finiteness and conformal invariance akin to \cite{Hull:1985rc} is left to the future as it requires a firmer understanding of how the sigma model sees the double geometry. The dilaton equation of motion of the double field theory should arise as an integrability condition of the generalised metric one (the dilaton beta functional occurs at higher look in $\alpha'$ so this is the easiest way to see it)~\cite{Callan:1986jb,Berman:2007yf}.   It would be interesting to consider whether things become clear upon using a derivative more related to the double differential geometry, like the semi-covariant one of \cite{Jeon:2011cn} whose connection contains the dilaton.

\subsection{The background field equation}

Given the result for the one-loop effective action (\ref{1loopresult}) we can find the beta-functional in exactly the same manner as for the ordinary string. The propagator integral at zero distance contained in $\Delta_0$ is divergent and can be regulated using dimensional regularisation, introducing a mass scale. In this simple case at one loop the beta-functional for $\H_{MN}$ is just proportional to the coefficient of $\d_1X^M\d_1X^N$\ in the effective action. That is, with original action
\bea
S^{cl}= \frac{1}{2} \int d^2\sigma\left[ -\H_{MN} \partial_1 X^M \partial_1 X^N + L_{MN} \partial_1 X^M \partial_0 X^N \right]\,   ,
\eea
and one-loop modification from (\ref{1loopresult})
\beq
S^{1\, loop}= \frac{1}{2} \int d^2\sigma\left[ -W_{MN} \partial_1 X^M \partial_1 X^N \right] =\int d^2\sigma\left[ \R_{MN} \partial_1 X^M \partial_1 X^N \right]\,   ,
\eeq
then the beta-function in proportional to $W_{MN}$ ($W$ was notation introduced in \cite{Berman:2007xn}). We confirm the conclusion of \cite{Copland:2011yh} that 
\beq
\R_{MN}=-\frac{1}{2}W_{MN}\, ,
\eeq
with no renormalisation of $L$. For the Weyl invariance of the sigma model to be preserved at one loop the beta-functional must vanish, thus so must $\R_{MN}$. This requirement of generalised Ricci flatness is the generalised metric equation of motion of the double field theory, justifying the statement that doubled field theory is the effective field theory of the more general sigma model. Since the double field theory equation of motion is equivalent to those from  the ordinary string effective action \eqref{EffAct} this also establishes the quantum equivalence of the double sigma model to the ordinary string case.

\section{Discussion}

We have shown that the double sigma model \eqref{NDaction} is classically Lorentz invariant and has double gauge symmetry. The vanishing of the one loop beta-functionals is equivalent to the generalised metric equation of motion of doubled field theory: the vanishing of the generalised Ricci tensor $\R_{MN}$. Just like the action (\ref{EffAct}) is the effective action for bosonic string sigma model (or massless NS-NS sector of the superstring), double field theory is the effective action for the new sigma model. 
As in the fibred case (that of the doubled formalism), since both the double field theory and the sigma model can be undoubled and shown equivalent to the ordinary string versions, this is not surprising, but quantum equivalence is not guaranteed. It is however reassuring, demonstrating the consistency of the double sigma model and double field theory. Moreover, it is through the vanishing of $\R_{MN}$ that they are connected, illustrating the central r\^ole of this tensor. It combines the equations of motion of the massless string fields in a manner revealing an underlying $O(D,D)$ symmetry and should arise as some kind of curvature on the double geometry.
The next step, which we will pursue in future work, is to calculate the two-loop beta functional and thus find the higher order corrections to the double field theory, which have been sought but have proven difficult to pin down (see for example \cite{Hohm:2011si}).

In the background field expansion one can choose to expand in any parameter, and we have chosen an expansion covariant with respect to the the Levi-Civita connection treating $\H$ as the metric. In this case it would be more natural to expand in an $O(D,D)$ connection which annihilates $L$ as well as $\H$ (perhaps like those featuring in \cite{Hohm:2010xe,Jeon:2011cn,Coimbra:2011nw}) but before this one must understand who one interprets things like geodesics in the doubled space. The correct expansion would be especially pertinent in attempting a higher loop calculation, and might even provide an efficient way of calculating the corrections to the undoubled effective theory, as the anti-symmetric tensor would also be included in $O(D,D)$ symmetric manner. Also, since the full fermionic part of double supergravity has not been finalised yet\cite{Hohm:2011zr,Jeon:2011vx,Coimbra:2011nw}, perhaps a supersymmetric sigma model could be of use.

The sigma model (\ref{NDaction}) is the analogue of the non-manifestly Lorentz invariance obtained after the PST procedure in \cite{Berman:2007xn}, rather than the manifestly Lorentz invariant action with chirality constraint of the doubled formalism. Another question is whether the manifestly Lorentz invariant action with constraint exists in the more general case. It should contain additional ingredients of the doubled formalism like connections for the fibre that we have not included here, when related to it by a reduction. Since only half the components of the equation of motion can be integrated and are needed, only they need be imposed by the Lagrange multipliers in the PST procedure, so it is unclear what an O(D,D) symmetric form of the constraint would look like. In \cite{Hull:2006va} when imposing the constraint by gauging currents it was also found only half the components had to be enforced. There is also some discussion of the general unfibred case in \cite{Hull:2006va} and the approach used there could be an alternative starting point in the search for such an action.

Also, the full version of the original, truly doubled, double field theory (with only the weak form of the level matching constraint imposed) has not been found. In fact recent work on compactifying double field theory to obtain gauged supergravites has shown that imposing the strong constraint does not lead to the most general gauging\cite{Hohm:2011si}, and that the double field action is still consistent and gauge invariant with some relaxation of the strong and even weak constraints\cite{Grana:2012rr} (a slight relaxation in supersymmetric extension to double field theory can lead to massive IIA supergravity\cite{Hohm:2011cp}).
As stated in the introduction such a relaxation in the sigma model would also be of interest. Perhaps also a group construction like in the sigma models of\cite{Sfetsos:2009vt,Avramis:2009xi} would allow the strong constraint to be relaxed whilst maintaining enough structure to be tractable.

As presented here both the sigma model and double field theory make no distinction between directions with or without isometry, but it is only in the former case that $O(D,D)$ rotations relate equivalent undoubled string backgrounds. In other directions only the canonical frame would have an interpretation as a geometric string background as those with non-trivial dual co-ordinate dependence would have to be interpreted as having windings round an infinite circle. Though not conventional string backgrounds, some of them may be those got by generalised T-duality in \cite{Dabholkar:2005ve}, which can be locally non-geometric (compared to T-fold backgrounds which are at least locally geometric) and could be related to geometric backgrounds by $O(D,D)$ rotations in directions without isometry in the double theories we consider. Since investigation of non-geometric sting backgrounds was one of the motivations of studying sigma models with doubled target spaces, we hope the sigma model \eqref{NDaction} and connection to double field theory could prove useful in this regard.

\section*{Acknowledgements}

I am very grateful to Jeong-Hyuck Park and Daniel Thompson for useful conversations and comments. This work was supported by the National Research Foundation of Korea (NRF) grant funded by the Korea government(MEST) with the Grant No. 2005- 0049409 (CQUeST).

\newpage

\newpage
%\section*{References}
\addcontentsline{toc}{chapter}{\sffamily\bfseries Bibliography}
\bibliographystyle{BibliographyStyleM}
\bibliography{DoubleBib.bib}
\end{document}